# Anomalous resistivity upturn in epitaxial $L2_1$-Co$_2$MnAl films

L. J. Zhu[1,2,*] & J. H. Zhao[2]

[1]*Cornell University, Ithaca, NY 14850, USA*

[2]*State Key Laboratory of Superlattices and Microstructures, Institute of Semiconductors, Chinese Academy of Sciences, P. O. Box 912, Beijing 100083, China*

* Correspondence should be addressed to L.J.Z. (zhulijun0@gmail.com)

**Despite of the great scientific and technology interest, highly ordered full-Heusler $L2_1$-Co$_2$MnAl films have remained a big challenge in terms of the availability and the electrical transport. Here we report the controllable growth and the intriguing transport behavior of epitaxial $L2_1$-Co$_2$MnAl films, which exhibit a low-temperature ($T$) resistivity upturn with a pronounced $T^{1/2}$ dependence, a robust independence of magnetic fields, and a close relevance to structural disorder. The resistivity upturn turns out to be qualitatively contradictory to weak localization, particle-particle channel electron-electron interaction (EEI), and orbital two-channel Kondo effect, leaving a three-dimensional particle-hole channel EEI the most likely physical source. Our result highlights a considerable tunability of the structural and electronic disorder of magnetic films by varying growth temperature, affording unprecedented insights into the origin of the resistivity upturn.**

The full-Heusler alloy $L2_1$-Co$_2$MnAl has attracted considerable attention in the emerging field of spintronics due to the vanishingly small density of states in minority spin band at Fermi level[1,2]. Firstly, Heusler ferromagnets with high spin polarization ($P$) have promise for amazingly high giant magnetoresistance (GMR) in current-perpendicular-plane (CPP) spin valves (SVs) and tunneling magnetoresistance in magnetic tunneling junctions (MTJs)[3]. Furthermore, the gap or pseudo-gap in the minority spin band significantly suppresses the spin-flip scattering and leads to a very small Gilbert damping constant[4,5] which is a predominant advantage for low-critical-current spin-torque magnetic memories[6] as well as long-distance spin wave transport.[7] Moreover, as proposed by J. Park *et al.*,[8] high-$P$ Heusler alloys could be an excellent approach to solve the over-rotation problem and to realize ultrafast (~100 ps) deterministic switching of the orthogonal spin torque (OST) CPP SVs. The $L2_1$-Co$_2$MnAl films are also of great interest in high-sensitivity Hall sensor application due to the giant intrinsic anomalous Hall conductivity arising from Berry phase curvature[1,2]. However, the control of the structural disorder is a crucial issue for the practical Co$_2$MnAl films in order to obtain the fascinating properties expected in perfectly ordered samples. As shown in Fig. 1(a), $L2_1$-Co$_2$MnAl crystallizes in the cubic space group $Fm\overline{3}m$ with Co, Mn and Al atoms occupying the Wyckoff position (1/4, 1/4, 1/4), (0, 0, 0), and (1/2, 1/2, 1/2), respectively. $L2_1$-Co$_2$MnAl is predicted to have a high $P$ of 75% and a magnetization ($M_s$) of ~4.04 $\mu_B$/f.u. (i.e. ~786 emu/cm$^3$)[1,2]. $L2_1$-Co$_2$MnAl bulk was obtained by induction melting and 10 day annealing at 600-1190ºC under a magnetic field of 5 kOe[9]. However, a $L2_1$-ordered Co$_2$MnAl thin film which is of interest for spintronics has barely been achieved despite the extensive efforts[3-5,10,11]. The Co$_2$MnAl films usually show a $A2$ or $B2$ ordering when epitaxially grown on semiconductors[3-5,10] or when sputtered on MgO (001) with/without a Cr buffer even after post-annealing at up to 500ºC[11]. Despite the Mn-Al swaps preserve the high $P$ and $M_s$[1,2], the Co-Mn swaps significantly reduces both $P$ and $M_s$ as a consequence of the antiparallel alignment of spin moments of Co-Mn or Mn-Mn atoms as revealed by first-principles calculations[13] and x-ray magnetic circular dichroism measurements[10]. A controllable growth of highly ordered $L2_1$-Co$_2$MnAl is, therefore, of great interest for understanding this material and its spintronic applications despite the technical challenge. On the other hand, the resistivity ($\rho_{xx}$) of $L2_1$-Co$_2$MnAl films is critical for both the magnitude and temperature ($T$) dependence of the GMR in CPP SVs because the GMR is expected to scale with $\rho_{xx}$ and spin-asymmetry coefficients according to Valet-Fert model[14]. The OST CPP SVs with a high-$P$ free layer of $L2_1$-Co$_2$MnAl are of great technological interest as sub-ns cryogenic cache memories that can be integrated with Josephson junction logics with 40-100 GHz clock frequency[8]. Therefore, it is of extreme importance to control the low-$T$ transport behaviors for a better SV performance, for instance, via taking advantage of the structural disorder effects of Co$_2$MnAl films. Structural disorder such as impurities, strain, and dislocations in Mn-based alloys has been proved to account for the large variations in magnetooptical Kerr effects[15,16], Gilbert damping[17], magnetic anisotropy[18], anomalous Hall effect[19], and for the occurrence of electronic disorder physics (e.g. weak localization[20] and atomic tunneling effects[21-23]). However, so far, the disorder effects on the transport properties of Co$_2$MnAl films have been barely investigated. Here we for the first time report an engineering of the structural and electronic disorder in epitaxial $L2_1$-Co$_2$MnAl films via controlling the growth temperatures. We observed an intriguing low-$T$ resistivity upturn with a $T^{1/2}$ scaling that exhibits close relevance to structural disorder and a robust independence of strong external magnetic fields ($H$). Such an anomalous resistivity upturn is clarified to arise from three-dimensional (3D) electron-electron interaction (EEI) in a particle-hole diffusion channel.

## Results

**Samples and spin polarization.** A series of 40 nm $Co_2MnAl$ films were grown on 150 nm GaAs-buffered semi-insulating GaAs (001)(see Fig. 1(b)). In order to tune the structural and electronic disorders of these $Co_2MnAl$ films, the substrate temperature ($T_s$) was varied to be 50, 150, 250, and 350 ºC, respectively. The composition was designed to be the same for different samples by controlling the fluxes from three thermal-diffusion cells and later verified to be Co: Mn: Al = 51 at%:17 at%: 32 at% by energy dispersive x-ray spectroscopy analysis. The epitaxial growth of the $Co_2MnAl$ films was confirmed by *in situ* reflective high energy electron diffraction (RHEED) patterns and *ex situ* synchrotron x-ray diffraction (XRD) patterns (see Fig. 1(c)). Each film was capped with a ~4 nm MgO protective layer at room temperature. The film thickness was confirmed by the oscillating x-ray reflectivity (XRR) curves at low incidence angles ($2\theta \leq 4º$) in Fig. 1(c). Notably, the oscillation magnitude varies significantly among different samples, indicating that the film homogeneity changes from pretty good at 50 ºC to relatively poor at 150 ºC, and then gradually improves with $T_s$ further increases to 350 ºC. From the XRD $\theta$-$2\theta$ pattern at high angles ($20º \leq 2\theta \leq 70º$), we can observe the (002) and (004) peaks of $Co_2MnAl$, indicating a superlattice of Co and (Mn,Al) atomic layers stacking along the perpendicular $c$ axis as schematically shown in Fig. 1(a). A small peak corresponding to the MgO (002) diffraction appears at $2\theta = 42.9º$ for each film, suggesting the epitaxial growth of the capping layer on $Co_2MnAl$ surface. For the sample with $T_s$=350 ºC, a weak peak unrelated to the $B2$ or $L2_1$ ordering appears at $2\theta = 33.55º$. As shown in Fig. 1(d), a fourfold (111) superlattice peak, the structural fingerprint of $L2_1$-ordering, was found for each $Co_2MnAl$ film in the high-resolution XRD $\varphi$ scans, indicating the alternatively stacking of Co, Mn, and Al monolayers. This is excellently consistent with the high-resolution tunneling electron microscopy measurements which showed clear $Co_2MnAl$ (111) diffraction spots in the selected area fast Fourier transform analysis when electron beam was parallel to $Co_2MnAl$ (110) direction[24]. Such excellent epitaxial compatibility with GaAs is fascinating in efficient pure spin injection into III-V semiconductors via spin pumping that is free of impedance problem and in the development of room temperature ferromagnetic semiconductors via magnetic proximity effect[25]. These films were patterned into Hall bars to measure the $T$ and $H$ dependences of $\rho_{xx}$ with a direct current flowing along $Co_2MnAl$ [110] (see Fig. 2(a)).

Since the spin polarization of conduction band at the Fermi level is crucial for spin-related electron scattering processes (e.g. $s$-$d$ electron scattering and two-channel Kondo (2CK) effect) of a ferromagnetic film and thus also the magnitudes and the $T$ profiles of the GMR in CPP SVs, we first try to shed some light on the bulk $P$ of the $L2_1$-$Co_2MnAl$ films through the possible indirect evidences, which will also benefit the understanding of the anomalous resistivity upturn later. First-principles calculations and experiments show that the Co-Al swaps for $Co_2MnAl$ alloy is energetically much more difficult to occur than Mn-Al and Co-Mn swaps[13]; the Mn-Al disorder has negligible influence on $P$ and $M_s$. There is a wide consensus that a high bulk $P$ in a $Co_2MnAl$ sample should be accompanied by a $M_s$ comparable with theoretical value[1,3,12,13], while the occurrence of many Co-Mn swaps would significantly reduce the $P$ and $M_s$ at the same time. The clear presence of $L2_1$-ordering in XRD results and the high $M_s$ for $T_s$ >150 ºC (see Fig. 2(d)) suggest few Co-Mn or Mn-Al swaps and thus a relatively high $P$ value[3]. The slightly higher $M_s$ than theoretical value may be probably due to the off-stoichiometry of these films[3,13]. The relatively small $M_s$ of 338 emu/cm³ at $T_s$ = 50 ºC indicates an increased Mn-Mn antiferromagnetic coupling[10,13] as the strong (002) and (111) XRD peaks make enhanced Co-Mn swaps unlikely. In addition, the bulk $P$ in a Heusler alloy may be indirectly indicated by the AMR which basically originates from $s$-$d$ scattering with localized $d$ states hybridized with opposite spin states by spin-orbit interaction. When one of the spin bands has very small density of states (e.g. $L2_1$-$Co_2MnAl$) or is even absent (e.g. half metal) at the Fermi level, the $s$-$d$ scattering occurs dominatingly in the other spin channel, which would favor a AMR that makes $\rho_{xx}$ larger when the current flows orthogonally to magnetization ($\rho_{xx}^{\perp}$) than when parallel ($\rho_{xx}^{\parallel}$)[26]. Figure 2(b) shows the magnetoresistance (MR) curves of the $L2_1$-$Co_2MnAl$ films measured at 300 K by sweeping a perpendicular magnetic field. For each film a small MR shows a dip at zero filed and gradually increases to peak values at ~ ±15 kOe due to the change of magnetic moment orientation with respect to current, i.e. AMR. These films have a magnetically easy axis along [110] direction and a hard axis along [001] direction as indicated by the magnetization hysteresis in Fig. 2(c). With $H$ gradually increasing from zero to beyond ±15 kOe[24], the magnetization gradually rotate from the parallel (i.e. [110]) to the orthogonal direction (i.e. [001]) with respect to the charge current, leading to a variation in measured $\rho_{xx}$ due to the AMR effect. Therefore, the small $\rho_{xx}$ at zero field compared with that at ~±15 kOe is consistent with a relatively high bulk $P$ of the high-quality $L2_1$-$Co_2MnAl$ epitaxial films. Intriguingly, we also observed a small MR in the $Co_2MnAl$ films at high $H$, which scales linearly with $H$ in a broad $T$ range from 2 to 300 K[24]. Though not yet fully understood, this high-$H$ MR should be irrelevant to AMR as it does not saturate even at 8 T but the magnetic moment aligns completely along film normal when $H$ is beyond ±15 kOe. Spin wave scattering can be excluded as thermal magnons are unlikely to be excited at $T$ as low as 2 K. Taking into account the evident presence of $L2_1$-ordering in XRD patterns, the large $M_s$, and the AMR with $\rho_{xx}^{\perp} > \rho_{xx}^{\parallel}$, we may expect in these $L2_1$-$Co_2MnAl$ films a $P$ that is at least comparable with sputtered $B2$ films ($P$ ~ 60%).[3]

Note that a half metallicity ($P\sim100\%$) should not be expected for $Co_2MnAl$ since it is not a half metal in theory ($P$ is theoretically no larger than 75%). We also mention that spin-polarized surface analysis techniques and $Al_2O_3$-barrier MTJs can only be indicative of a surface/interface spin polarization instead of the bulk $P$.

**Temperature dependence of the longitudinal resistivity.** Now we turn to the $T$-dependent electrical transport behaviors of these films. Figure 3(a) shows the zero-field $\rho_{xx}$ of the $Co_2MnAl$ films as a function of $T$. Each film shows a resistivity minimum ($\rho_{xx0}$), beyond which $\rho_{xx}$ increases monotonically with $T$ mainly due to the increasing phonon scattering. The apparent nonlinearity of $\rho_{xx}$-$T$ curve at high $T$ should be attributed to a weak magnon scattering, which further evidences that $L2_1$-$Co_2MnAl$ is not a complete half metal. In the following, we show that the low-$T$ resistivity upturns in these $Co_2MnAl$ films most likely arise from disorder-enhanced EEI in a particle-hole channel. The 3D particle-hole channel EEI is expected to give a $T^{1/2}$-dependent correction to $\rho_{xx}$ as a consequence of the correlation between wave function of the added electron and the wave functions of the occupied electrons that are nearby in energy[27]. In the presence of a nonzero $H$, the correction includes a $S_z=0$ Hartree contribution and a $S_z=\pm1$ triplet contribution. The former involves electrons with the same spin, and is unaffected by the splitting of spin up and down bands; while the latter should be independent of $H$ in a ferromagnetic system because both Zeeman splitting (~0.68 meV at 6 T) and the $T$ (4.3 meV at 50 K) are negligible compared with the ferromagnetic exchange splitting (~10 eV). Therefore, 3D particle-hole channel EEI in a ferromagnet can be signified by a $T^{1/2}$-dependent resistivity upturn with an $H$ independence and a close correlation to disorder.

Figures 3(b) plots the resistivity variation at $H = 0$ T as a function of $T$ for the $Co_2MnAl$ films with different $T_s$. The resistivity increase, $\Delta\rho_{xx}$ ($\Delta\rho_{xx} = \rho_{xx}-\rho_1$, with the offset $\rho_1$ determined from the best linear fit of $\rho_{xx}$-$T^{1/2}$), varies linearly with $T^{1/2}$ when $T$ drops below a typical temperature $T_0$. $T_0$ ($\rho_1$) was estimated to be ~15.2 (128.1), ~27.8 (155.0), ~30.0 (151.0), ~17.9 (156.1) K ($\mu\Omega$ cm) for $T_s$ = 50, 150, 250, and 350 $^o$C, respectively. The slope $k=d\rho_{xx}/d(T^{1/2})$ reflects the strength of the EEI and the diffusion constant of the film, i.e. an increase in $k$ suggests an enhancement of the EEI and/or a reduction of diffusion constant[27]. The $T^{1/2}$ dependence of the resistivity upturns in different samples can be visualized more directly by collapsing the $T$-dependent $\Delta\rho_{xx}$ data of different samples onto single scaling curve of $-\Delta\rho_{xx}/kT_0^{1/2}= (T/T_0)^{1/2}$ for $T/T_0<1$. From Fig.3(c) we can see that $\log(-\Delta\rho_{xx}/kT_K^{1/2})$ scales linearly with $\log(T/T_K)$ with a slope of 1/2 when $T/T_0 < 1$ for the $Co_2MnAl$ samples with different $T_s$. It should be mentioned that the high-field MR which even exists at 300 K (see Fig. 2(b)) is unlikely to be related to the $T^{1/2}$ dependence of $\Delta\rho_{xx}$ due to the EEI.

**Magnetic field independence and structural disorder relevance.** In order to establish more rigorously the particle-hole channel EEI in these $Co_2MnAl$ films, we examined the effect of $H$ on the $T$ dependence of $\rho_{xx}$. As an example, we show $\Delta\rho_{xx}$ for $T_s = 150$ $^o$C under various $H$ of 0, 3, and 6 T in Fig. 3(d). The magnetic fields show no measurable effect on the $T$ dependence: $\rho_{xx}$ scales linearly with $T^{1/2}$ at below $T_0$ ($T_0$ ~27.8 K) and $T_0$ and $k$ are apparently independent of $H$, strongly suggesting a spin-independent origin of the resistivity upturns. The same features hold for other films with different $T_s$. As plotted in Fig. 3(e), $k$ first increases from the minimum of 0.142 $\mu\Omega$ cm/$K^{1/2}$ at 50 $^o$C to the maximum of 0.182 $\mu\Omega$ cm/$K^{1/2}$ at 150 $^o$C, and finally goes down to 0.169 $\mu\Omega$ cm/$K^{1/2}$ at 350 $^o$C, which reflects the evolution of the EEI strength and the electron diffusion constant with structural disorder[27]. The $T_s$ dependence of $k$ in the $Co_2MnAl$ films excellently agrees with those of $M_s$ and film homogeneity indicated by the XRR oscillation amplitude. We now discuss the evolution of the structural disorder with $T_s$ in these films and their close relevance to the resistivity upturn. Figure 4(a) summarizes as a function of $T_s$ the lattice constant $c$ of the $Co_2MnAl$ films determined from the $Co_2MnAl$ (004) XRD peaks. As schematically shown, their much larger $c$ than the bulk value of 5.755 Å[9] indicate a strong strain in these films probably due to the large lattice mismatch with GaAs ($a_{GaAs}$=5.662 Å), which also makes these films likely to have high-density dislocations. $c$ firstly drops from 5.880 Å at 50 $^o$C to 5.810 Å at 150 $^o$C, and then monotonically goes up up to 5.850 Å at 350 $^o$C, suggesting a significant variation in the strain strength. Compared to other samples, the slightly short $c$ axis at 150 $^o$C likely indicates a relatively high degree of strain relaxation and a high density of the dislocations. The evolution of the film disorder with $T_s$ can be further proved by the intensity of XRD peaks. Figure 4(b) shows the integrated intensity of $Co_2MnAl$ (004) peaks ($I_{004}$) normalized by the real-time monitored and recorded intensity of the incidence x-ray beam during the measurements. $I_{004}$ first drops remarkably by a factor of 10 as $T_s$ increases from 50 to 150 $^o$C, and then gradually increases as $T_s$ further increases towards 350 $^o$C, the tendency of which excellently coincides with those of homogeneity and strain. Intriguingly, in $L1_0$-MnGa and $L1_0$-MnAl films[19,21,22], if we use $\rho_{xx0}$ (approximately the residual resistivity due to static impurity scattering) and the thermal contribution to the resistivity ($\rho_{them}=\rho_{xx}-\rho_{xx0}$) to quantify the strengths of structural disorder and the phonon and magnon scattering, respectively, the phonon and magnon scattering can be found to become increasingly suppressed with enhancing structural disorder. Therefore, $\rho_{them}$ also reflects the degree of structural perfection. The same is also true for these $Co_2MnAl$ films as $\rho_{them}$ in Fig. 4(c) varies with $T_s$ in an analogue manner of strain and $I_{004}$. We also plotted $\rho_{xx0}$ as a function of $T_s$ in Fig. 4(d), which leads to the virtually same conclusion on how the structural disorder varies with $T_s$ except that the slight recovery of $\rho_{xx0}$ at $T_s$=350 °C in comparison to $T_s$=250 °C may be related to the appearance of the peak at $2\theta = 33.55^o$ and seems to be less important to the EEI. In a word, the $T^{1/2}$ scaling, the $H$-independence, and the disorder nature of low-$T$ resistivity upturns are highly consistent with a particle-hole channel EEI mechanism.

**Origin of the low temperature resistivity upturn.** We further clarify that particle-hole channel EEI is the most likely physical source of the low-$T$ resistivity upturns. Besides the particle-hole channel EEI, there are at least three alternative effects that can give rise to a $T^{1/2}$-dependent resistivity upturn: (i) 3D weak localization, (ii) 3D EEI in a particle-particle channel, and (iii) orbital 2CK effect due to resonant two-level system (TLS) scattering. An explanation of quantum correction due to 3D weak localization can be safely excluded here as the strong external magnetic field $H$ and the giant internal magnetic field due to ferromagnetic exchange splitting of ~$10^5$ T can completely destroy the quantum interference phase of the electron wave function on the time-reversed path and thus forbid the weak localization[27]. The 3D EEI in a particle-particle channel is expected to give a $T^{1/2}$-dependent correction to the resistivity at very weak magnetic field, however, a strong magnetic field can also easily destroy it. A $T^{1/2}$-dependent resistivity upturn is also expected in an orbital 2CK effect which occurs when a pseudospin-1/2 of structural TLS (where an atom or atom group with small effective mass coherently tunnels between two nearby positions) equally couples to two spin channels of conduction electrons via resonant scattering[28]. The orbital 2CK effect is manifested in electrical transport by a unique low-$T$ resistivity upturn which scales with $\ln T$, $T^{1/2}$ and $T^2$ in three distinct $T$ regimes, respectively[21,23]. Importantly, an external magnetic field should have no sizeable influence on the resistivity upturn because the electron spin variables are irrelevant to the Kondo coupling between the TLS and conduction electrons, and Zeeman splitting is negligible in comparison to the Fermi energy. The orbital 2CK fixed point and the resultant $T^{1/2}$ dependence also appear to be robust against a weak population asymmetry of the two spin channels (e.g. $L1_0$-MnGa and $L1_0$-MnAl films with an estimated bulk $P$ of <12.5%)[21-23]. However, the $P$ appears not to be low in these $L2_1$-Co$_2$MnAl films ($P$~60%-75%) as discussed above. The resultant strong channel asymmetry would lead to different tunneling rates of a TLS for two spin channels and decouple the TLSs from one spin channel, and consequently destroy the 2CK physics and the $T^{1/2}$ dependence of resistivity. Therefore, the 2CK effect seems unlikely here despite the agreements in the disorder dependence, the $T^{1/2}$ scaling, and the $H$-independence of resistivity upturn. On the other hand, in contrast to the $L1_0$-MnGa and $L1_0$-MnAl films displaying the orbital 2CK effect[21,23], these $L2_1$-Co$_2$MnAl films show a clear absence of a $\ln T$ dependence in the resistivity at higher energy scale, which further implies the inapplicability of 2CK explanation here. Here, we mention that our systematic study on Co$_2$MnAl films with controlled disorder may also suggest a diffusion channel EEI mechanism for the low-$T$ resistivity upturns in Heusler Co$_2$Mn$_{0.25}$Ti$_{0.75}$Al [20], Co$_2$MnSi [29], and Co$_2$MnGa films [30], while the weak localization explanation in the literature seems problematic.

**Discussions**

We have presented the controllable epitaxial growth and the transport behavior of $L2_1$-Co$_2$MnAl films. The $T^{1/2}$ dependence, the $H$ independence, and the disorder relevance of the low-$T$ resistivity upturn are well consistent with the 3D particle-hole channel EEI, while the weak localization, the particle-particle channel EEI and the orbital 2CK physics are conclusively excluded. The unprecedented tunability of the structural and electronic disorders leads to the insight into the origin of the resistivity upturn. The controllable epitaxial growth of highly ordered $L2_1$-Co$_2$MnAl films shows promise for spin pumping spin injection into III-V semiconductors and for developing room temperature ferromagnetic semiconductors utilizing magnetic proximity effect. These results would also benefit the understanding of the transport properties of various Heusler films and their spintronic applications such as ultrafast reliable cryogenic OST CPP-SV memories.

**Methods:**

**Sample preparation and characterizations.** The samples were prepared by a VG-80 molecular-beam epitaxy system with two growth chambers (one for III-V group semiconductors, the other for metals). For each film, a semi-insulating GaAs (001) substrate was first loaded into the semiconductor chamber to remove the oxidized surface by heating up to 580 °C in Arsenic atmosphere (~1×10⁻⁷ mbar) and to get a smooth fresh surface by growing a 150 nm GaAs buffer layer. Afterwards, the sample was transferred to second growth chamber to grow the 40 nm thick Co$_2$MnAl film at a rate of ~1 nm/min at 200 °C and a 4 nm thick MgO protective layer at room temperature. The film composition was determined by a Bruker energy dispersive x-ray spectroscopy with electron energy of 20 keV and working distance of 9.5 mm. The structure was measured by a reflective high energy electron diffraction, a synchrotron x-ray diffractometer at 4B9A beamline of Beijing Synchrotron Radiation Facility (BSRF), and a high-resolution Rigaku Smartlab x-ray diffractomerter with a Ge (220)×2 monochromator, respectively. The magnetism was measured by a Quantum Design superconducting quantum interference device magnetometer at room temperature.

**Devices fabrication and transport measurement.** The film was patterned into 60 µm wide Hall bars with an adjacent electrode distance of 200 µm using ultraviolet photolithography and ion-beam etching for transport measurements. The longitudinal resistivity ($\rho_{xx}$) were measured in a Quantum Design physical property measurement system as a function of the temperature and the perpendicular magnetic field with a 10 µA excitation current.

at 2 K, and out-of-plane magnetization hysteresis for $L2_1$-Co$_2$MnAl films.

## Acknowledgements

We gratefully acknowledge Q. Cai from BSRF and M. S. Weathers from Cornell Center for Materials Research for the help on synchrotron XRR and XRD scan and XRD $\varphi$ scan measurements, respectively. This work was partly supported by MOST of China (Grant No. 2015CB921503), NSFC (Grant No. 61334006), and NSF MRSEC program (Grant No. DMR-1120296).

## Author contributions

L. J. Z. designed and performed the experiments, L. J. Z. and J. H. Z. analyzed the data and wrote the manuscript.

## Additional Information

**Competing financial interests:** The authors declare no competing financial interests.

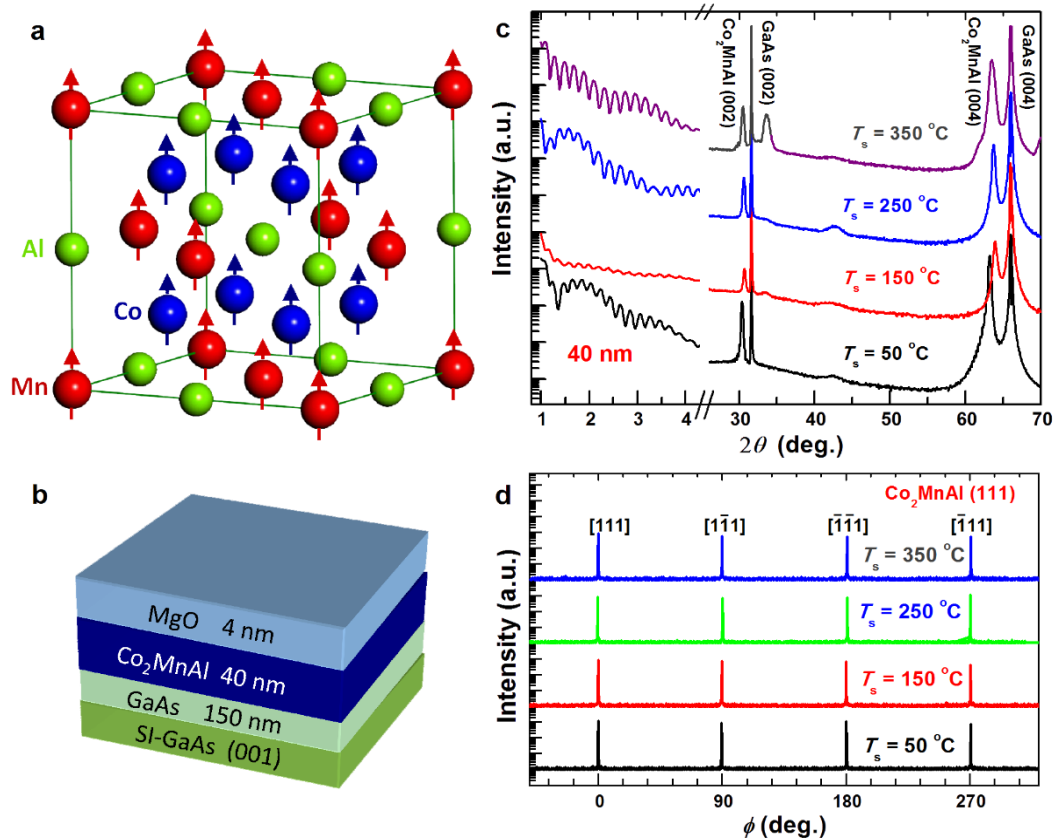

**Figure 1. Crystalline structure**. Schematic of (a) $L2_1$–Co$_2$MnAl lattice structure and (b) Sample structure; (c) XRR and XRD patterns and (d) $\varphi$ scan profiles for (111) peaks of the Co$_2$MnAl films grown at different $T_s$.

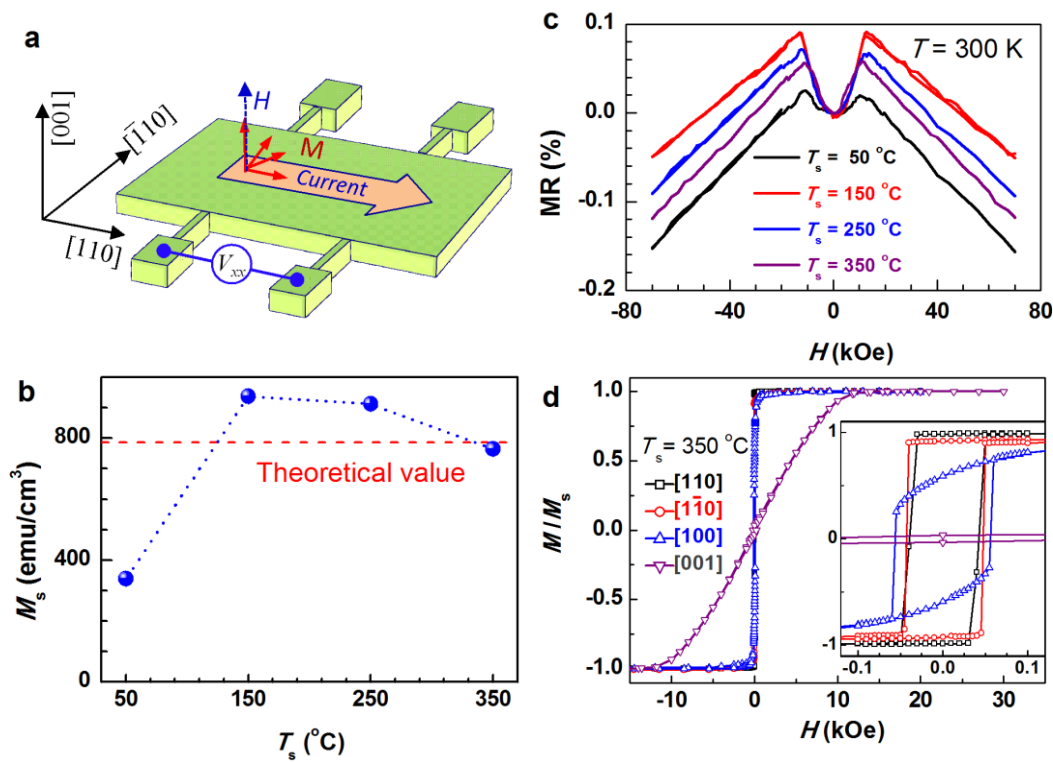

**Figure 2. Magnetism and magnetoresistance**. (a) Hall bar geometry with respect to the crystalline directions of Co$_2$MnAl; (b) $M_s$ and (c) MR at 300 K for the $L2_1$-Co$_2$MnAl film with different $T_s$; (d) Normalized magnetization hysteresis at 300 K ($T_s$ =350 ºC) along [110], [1$\bar{1}$0], [100] and [001] directions, respectively, the inset is close-up view at small fields. The red dashed line in (b) refers to the theoretical value of $M_s$.

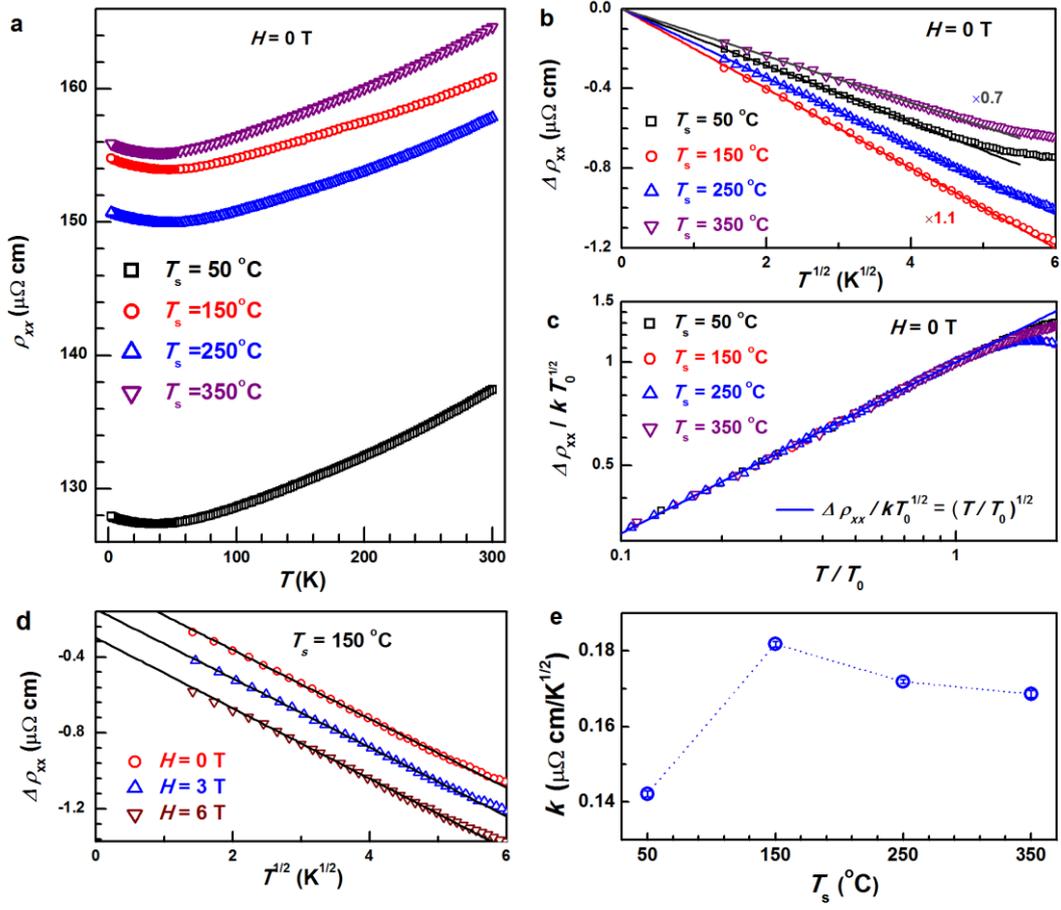

**Figure 3. *T* and *H* dependences of the longitudinal resistivity.** (a) *T* vs $\rho_{xx}$, (b) $\Delta\rho_{xx}$ vs $T^{1/2}$ ($H = 0$ T), (c) log-log plot of -$\Delta\rho_{xx}/kT_0^{1/2}$ vs $T/T_0$ ($H = 0$ T), (d) $\Delta\rho_{xx}$ vs $T^{1/2}$ ($H = 0$, 3, and 6 T), and (e) *k* for $Co_2MnAl$ films with different $T_s$. For clarity, $\Delta\rho_{xx}$ in (b) is multiplied by a factor of 0.7 and 1.1 for $T_s$= 50 and 150 ºC, respectively; the curves at 3 and 6 T in (d) were artificially shifted by -0.16 and -0.32 µΩ cm, respectively.

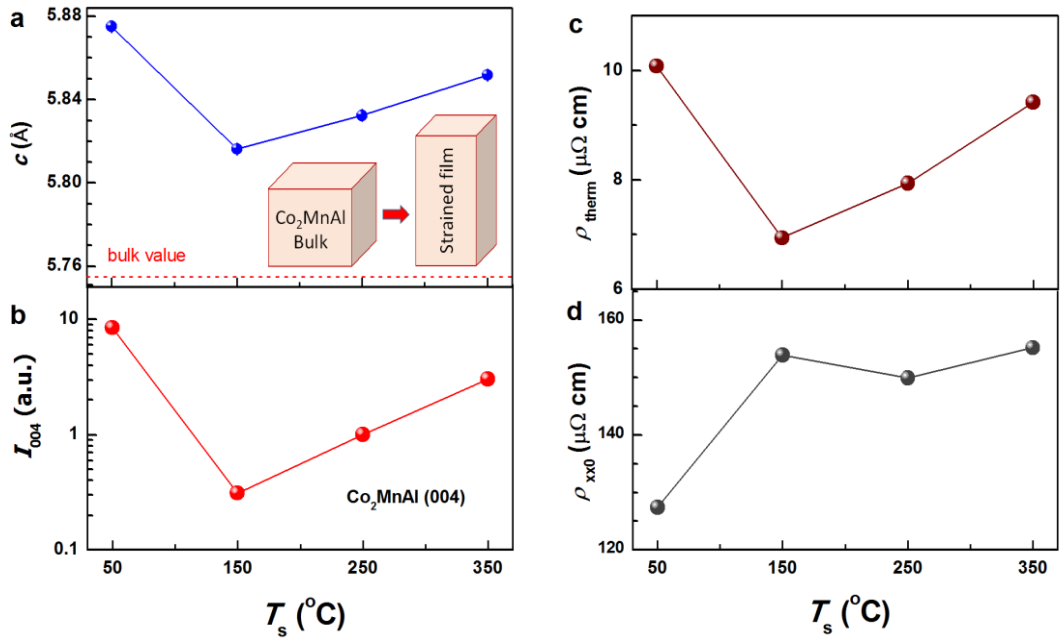

**Figure 4. *T_s* dependence of the structural and electronic disorder**. (a) *c*, (b) $I_{004}$, (c) $\rho_{therm}$, and (d) $\rho_{xx0}$ for $Co_2MnAl$ films with different $T_s$. The inset in (a) shows schematically a strained $Co_2MnAl$ lattice cell and a bulk one; the red dashed line shows the bulk value of *c*.